\documentstyle[aps,prl]{revtex}

\begin{document}
\title{Electron Magnetic Resonance: The Modified Bloch Equation}
\author{Bilha Segev$^{\,1}$ and Y.~B.~Band$^{\,1,2}$}
\address{${}^{1}$ Department of Chemistry, Ben-Gurion University \\
of the Negev, POB 653, Beer-Sheva 84105, ISRAEL \\
${}^{2}$ Atomic Physics Division, A267 Physics, \\
National Institute of Standards and Technology, Gaithersburg, MD 20899}
\maketitle

\begin{abstract}
We find a modified Bloch equation for the electronic magnetic moment
when the magnetic moment explicitly contains a diamagnetic
contribution (a magnetic field induced magnetic moment arising from
the electronic orbital angular momentum) in addition to the intrinsic
magnetic moment of the electron.  The modified Bloch is coupled to
equations of motion for the position and momentum operators.  In the
presence of static and time varying magnetic field components, the
magnetic moment oscillates out of phase with the magnetic field and
power is absorbed by virtue of the magnetic field induced magnetic
moment, even in the absence of coupling to the environment.  We
explicitly work out the spectrum and absorption for the case of a $p$
state electron.
\end{abstract}


\section{Introduction} \label{introduction}

The importance of magnetic resonance methods to chemical analysis,
solid-state physics, biological structure determination and medicine
is well recognized.  In magnetic resonance experiments, a material is
placed in dc and oscillating magnetic fields and the power absorption
is measured.  The power absorption exhibits spectra containing
resonances, from which valuable information about the system is
obtained.

The Bloch equation describing the dynamics of the magnetic moment
operator $\vec{\mu}$ is:
\begin{equation}
\frac{d\vec{\mu}}{dt} = -\frac{\mu_0 \gamma}{\hbar} \vec{B} \times
{\vec{\mu}} \, . \label{Blocheq}
\end{equation}
Its solution is discussed in textbook theoretical treatments of
magnetic resonance \cite{Textbooks}.  Here, $\mu_0 =
\frac{e\hbar}{2m}$ is the Bohr magneton, and $\gamma$ is a
dimensionless constant.  SI units are employed throughout.  This
equation has played a major role in understanding magnetic resonance
experiments ever since it was first derived by Bloch in 1946 \cite
{Bloch}.  In the usual treatment, coupling to the surrounding, i.e.,
to additional degrees of freedom of the system (often described as a
thermodynamic bath), is incorporated into the description by adding
phenomenological relaxation (decay) terms to the Bloch equation
\cite{Textbooks,Kittel,Kubo,Poole}.  The additional phenomenological
decay terms arise when the bath degrees of freedom of the system are
adiabatically eliminated from the description.  The external magnetic
field in magnetic resonance experiments has a temporally-constant
strong component, and a weaker oscillating component of
radio-frequency or microwave radiation, which in one common
configuration is linearly polarized and perpendicular to the dc
component:
\begin{eqnarray}
\vec{B}= B_0 \hat{z} + B_1 \left(e^{i \omega t}+e^{-i \omega t}\right)
\hat{x} \ ; \ B_1 \ll B_0 \ .  \label{B}
\end{eqnarray}
It is a straightforward generalization to treat the case when the
magnetic field contains a rotating oscillating field: $\vec{B}_{{\rm
c}} = B_0 \hat{z} + \sqrt{2} B_1 [\hat{x} \cos(\omega t) - \hat{y}
\sin(\omega t)]$.  Here we consider linear polarization.  For atomic
and molecular systems it is an excellent approximation to neglect the
spatial dependence of the oscillating field because the wavelength is
so much larger than the size of the atom or molecule.

The power absorption $P$ of a magnetic moment $\vec{\mu}$ in the
external field $\vec{B}$ is given by:
\begin{equation}
P=\left[ -\langle \vec{\mu}\rangle \cdot \frac{d\vec{B}}{dt}\right]
_{t} = \left[ \vec{B}\cdot \frac{d\langle \vec{\mu}\rangle}{dt}\right]
_{t}\ ,
\label{pap}
\end{equation}
where $\langle ...\rangle $ stands for quantum expectation value and
$\left[ ...\right] _{t}$ for time averaging.  Power can only be
absorbed if $\vec{B}$ and $\vec{\mu}$ oscillate out of phase, and
$\vec{\mu}$ has a component along $\frac{d\vec{B}}{dt}$.  The Bloch
equation without relaxation terms, Eq.~(\ref{Blocheq}), results
in no power absorption, since clearly, $\vec{B}\cdot \left(
\vec{B}\times \vec{\mu}\right) =0$.  In this treatment, the phase
shift between the oscillations of $\vec{B}$ and $\vec{\mu}$, necessary
for power absorption, exists only because of the relaxation terms. 
Without those, Eq.~(\ref{Blocheq}), implies that power absorption is
impossible.

This theoretical framework is adequate for the description of both NMR
(Nuclear Magnetic Resonance) and ESR (Electron Spin resonance).  It is
generally not correct for EMR (Electron Magnetic Resonance).  The Bloch
equation is not the correct equation of motion for the magnetic moment
operator of an electron with non-vanishing orbital angular momentum. 
Orbital angular momentum effects in magnetic resonance phenomena can be
important, not just in gases, but even in solids, e.g., the transition and
lanthanide ions doped in crystals where quenching (the modification of the
magnetic susceptibility to values close to those produced by electron spin
alone due to large crystal field splittings) does not occur
\cite{unquenched}.  Here we show that, due to diamagnetic effects arising
from the electronic orbital angular momentum, the Bloch equations are
replaced by a set of coupled operator equations for the magnetic moment
$\vec{\mu}$ (which incorporates a magnetic field induced magnetic moment
contribution), the coordinate $\vec{r}$ and the velocity operator
$\vec{\Pi}/m$, defined below.  The interaction Hamiltonian has non-linear
terms arising from a magnetic field induced magnetic moment. As a result,
the magnetic moment has an oscillatory component along
$\frac{d\vec{B}}{dt}$ and oscillates out of phase with the external field.
Hence, power absorption does not vanish, even when coupling to the
environment is not included. 

In Sec.~\ref{secH} we define the magnetic moment operator, including
the magnetic field induced magnetic moment contribution, and in
Sec.~\ref{Heq} we derive the Heisenberg equations of motion for a
magnetic moment $\vec{\mu} $ in the magnetic field $\vec{B}$.  Three
coupled equations are obtained in this way, Eqs.~(\ref{dmu}),
(\ref{dr}), and (\ref{dp}) below.  Based on these equations,
Sec.~\ref{power} develops general considerations of power absorption
in EMR experiments.  Since, unlike the Bloch equation, these coupled
equations can not be solved in closed form, we solve them in
time-dependent perturbation theory, and calculate the
power absorption using this analysis.  In Sec.~\ref{Example} we focus
on the example of an electron bound in a
$p$-atomic-orbital, and derive Eq.~(\ref {final}) for the power
absorption to leading order in the fields for this case. 
Section~\ref{Conclusions} presents a summary and conclusion.

\section{The Hamiltonian and the Magnetic Moment Operator} \label{secH}

The Hamiltonian for an electron, with charge $q=-e$ and mass $m$ in a
spherically symmetric potential $\hat{V}(r)$ and an external magnetic
field $\vec{B}(t)$, can be written in terms of the kinetic velocity
operator $\hat{\vec{\Pi}}/m$ \cite{Sakurai}:
\begin{eqnarray}
&&\hat{H}=\hat{V}(r)+\frac{\hat{\Pi}^2}{2m} \ ,  \label{H} \\
&&\hat{\vec{\Pi}}\equiv \hat{\vec{p}}-{q} \hat{\vec{A}} \ , \\
&&\hat{\vec{A}}\equiv \frac{1}{2} \vec{B}\times\hat{\vec{r}} \ .
\end{eqnarray}
For simplicity, we have neglected the electron spin (which can be
re-introduced rather easily as shown in Sec.~\ref{Conclusions}) and
chosen a symmetric gauge. The coordinate operator $\hat{\vec{r}}$,
the conjugate momentum operator $\hat{\vec{p}}$, and the velocity
operator $\hat{\vec{\Pi}} /m$, satisfy the commutation relations:
$[r_i,p_j]=[r_i,\Pi_j]=i \hbar \delta_{ij} $, and $[{\Pi}_i,\Pi_j]=i
\hbar q \epsilon_{ijk} B_k$, where $ \epsilon_{ijk}$ is the
anti-symmetric Levi-Chivita tensor.

The gauge-invariant operator for the orbital magnetic dipole moment of the
electron is given by 
\begin{eqnarray} 
&&\hat{\vec{\mu}}=\frac{q}{2m} \
\hat{\vec{r}}\times\hat{\vec{\Pi}} \ .  \label{mag_moment}
 \end{eqnarray}
The magnetic moment operator defined in Eq.~(\ref{mag_moment}) is gauge
invariant.  This can be seen by considering the gauge transformation
matrix $U = \exp[i\frac{q}{\hbar}\chi]$ in terms of the gauge function
$\chi(\vec{r},t)$.  Under the gauge transformation, the vector potential
and the magnetic moment are transformed as follows:  $\vec{A} \rightarrow
\vec{A} ^{\prime}= \vec{A} + \nabla \chi$, and $\hat{\vec{\mu}}
\rightarrow \hat{\vec{\mu}}^{\prime}= U[\frac{q}{2m} \
\hat{\vec{r}}\times(\hat{\vec{p}}-{q} \hat{\vec{A}^{\prime}})]U^{\dagger}
= \hat{\vec{\mu}}$. 

The Hamiltonian of Eq.~(\ref{H}) can be written in terms of the magnetic
moment in the following form: 
\begin{eqnarray}
&&\hat{H}=\hat{V}(r)+\frac{\hat{p}^2}{2m} -\hat{\vec{\mu}}\cdot\vec{B}
+ \frac{q^2}{8 m} \left[ (\vec{B}\cdot\hat{\vec{r}})^2 -B^2\hat{r^2}
\right] \ .  \label{Hmu}
\end{eqnarray}
Neglecting all terms of order $B^2$ gives:
$\hat{H}^{\prime}=\hat{V}(r)+ \frac{\hat{p}^2}{2m} - \hat{\vec{\mu
^{\prime}}}\cdot\vec{B}$ and $\hat{\vec{\mu ^{\prime}}} \equiv
\frac{q}{2m} \ \hat{\vec{r}}\times\hat{\vec{p}}$, as used in standard
textbook treatments of magnetic resonance.  Here, we consider the
effect of not neglecting these second order terms, i.e., using
$\hat{H}$ (not $\hat{H}^{\prime}$) with the full $\hat{\vec{\mu}}$
(not $\hat{ \vec{\mu ^{\prime}}}$).

\section{Heisenberg equations} \label{Heq}

The magnetic moment $\hat{\vec{\mu}}$ satisfies the Heisenberg
equation of motion: $\frac{d\hat{\vec{\mu}}}{dt}=\frac{\partial
\hat{\vec{\mu}}}{ \partial t}+ \frac{i}{\hbar }\left[
\hat{H},\hat{\vec{\mu}}\right]$.  Straight-forward algebra gives,
\begin{equation}
\frac{\partial \hat{\vec{\mu}}}{\partial t}=\frac{q}{2m}\
\hat{\vec{r}} \times \frac{\partial }{\partial t}\hat{\vec{\Pi}} =
-\frac{q^{2}}{4m}\ \hat{\vec{r}}\times \left( \frac{\partial
}{\partial t}\vec{B}\times \hat{\vec{r}} \right)
=\frac{q^{2}}{4m}\left[ \left( \hat{\vec{r}}\cdot \frac{\partial {\
\vec{B}}}{\partial t}\right) \ \hat{\vec{r}}\ -\
\hat{r^{2}}\frac{\partial {\ \vec{B}}}{\partial t}\right] \ ,
\end{equation}
\begin{eqnarray}
-\frac{i}{\hbar }\left[ \hat{H},\hat{\vec{\mu}}\right]
&=&\frac{q^{2}}{ 2m^{2}c^{2}}\left\{ (\hat{\vec{r}}\cdot
\hat{\vec{p}})\vec{B} - (\hat{\vec{r} } \cdot \vec{B})\hat{\vec{p}} +
\frac{q}{2c}\left( \hat{\vec{r}}\cdot \vec{B} \right) \left(
\vec{B}\times \hat{\vec{r}}\right) -i\hbar \vec{B}\right\} \nonumber
\\
&=&\frac{q^{2}}{2m^{2}c^{2}}\left\{ \hat{\vec{r}}\times (\vec{B}\times
\hat{ \vec{\Pi}})-i\hbar \vec{B}\right\} \ ,
\end{eqnarray}
and, collecting terms, we get: 
\begin{equation}
\frac{d\hat{\vec{\mu}}}{dt}=\frac{q^{2}}{4m}\left[ \left( \hat{\vec{r}}\cdot 
\frac{\partial {\vec{B}}}{\partial t}\right) \ \hat{\vec{r}}\ -\ \hat{r^{2}} 
\frac{\partial {\vec{B}}}{\partial t}\right] +\frac{q^{2}}{2m^{2}}\left[ 
\hat{\vec{r}}\times (\vec{B}\times \hat{\vec{\Pi}})-i\hbar \vec{B}\right] \ .
\label{dmu}
\end{equation}
The right hand side of Eq.~(\ref{dmu}) is not equal to
$\frac{\mu_0}{\hbar} \vec{B}\times{\vec{\mu}} \equiv \frac{q^2}{2 m^2}
\left[ \vec{B} \times ( {\vec{r}}\times \hat{\vec{\Pi}}) \right]$. 
The Bloch equations are replaced by operator equations whose
consistent solution must treat the Heisenberg equation of motions for
$\hat{\vec{\mu}}(t)$ as well as $\hat{\vec{r}}(t)$ and
$\hat{\vec{\Pi}} (t)$:
\begin{eqnarray}
&&\frac{d\hat{\vec{r}}}{dt} = \hat{\vec{\Pi}}/m \ ,  \label{dr} \\
&&\frac{d\hat{\vec{\Pi}}}{dt} =
\frac{\partial\hat{\vec{\Pi}}}{\partial t}- \frac{i}{\hbar} \left[
\hat{\vec{\Pi}}, V(\hat{r}) \right] - \frac{i}{\hbar} \left[
\hat{\vec{\Pi}}, \hat{\vec{\Pi}}^2/2m \right] \ .  \label{dp}
\end{eqnarray}

From these equations we see that energy can be absorbed by the system
even under conditions of negligible or no coupling to the environment
since ${d\vec{\mu}}/{dt}$ is no longer perpendicular to $\vec{B}$. 
{\em This power absorption in magnetic resonance systems containing
electronic orbital angular momentum can also be interpreted as being
due to the rate of change of the electric dipole moment in the
presence of an electric field}.  This is evident from the relation:
\begin{equation}
\frac{\partial \hat{H}}{\partial t}=\frac{e}{2m}\left(
\frac{d\vec{B}}{dt} \times \vec{r}\right) \cdot
\vec{\Pi}=-\vec{\mu}\cdot \frac{d\vec{B}}{dt}= \vec{E}\cdot
\frac{d\vec{\mu _{E}}}{dt}\ . \label{e_interp}
\end{equation}
Here the electron electric dipole moment is
$\vec{\mu}_{E}=q\vec{r}=-e\vec{r} $, its time derivative is
${d\vec{\mu}_{E}}/{dt}=-{e}\vec{\Pi}/m$, and the electric field is
given by $\vec{E}=-{\partial \vec{A}}/{\partial t}$.  Clearly,
additional power aborption may result due to coupling to a bath.

The time derivative of the magnetic dipole moment in Eq.~(\ref{dmu})
can not be written simply in terms of the cross product of the
external field and the dipole moment, hence a closed-form analytic
solution of the operator equations in the Heisenberg representation is
not available.  Instead a perturbative solution can be used, as
described below.

\section{Power absorption} \label{power}

Substituting the magnetic field of Eq.~(\ref{B}) into the Hamiltonian in
Eq.~(\ref{Hmu}), we find that the Hamiltonian can be separated into a sum of
time-independent and time-dependent parts: 
\begin{eqnarray}
&&\hat{H}= \hat{H}_0+\hat{H}_1(t) \ , \\
&&\hat{H}_0 =\hat{V}(r)+\frac{\hat{p}^2}{2m} -B_0 \frac{q}{2m}
(\hat{x}\hat{p }_y -\hat{y}\hat{p}_x ) + B_0^2 \frac{q^2}{8m}
(x^2+y^2)\ , \\
&&\hat{H}_1(t) = - B_1\left(e^{i \omega t}+e^{-i \omega t}\right)
\hat{h}_1 \ ; \ \ \hat{h}_1 \equiv \frac{q}{2m}(\hat{y}\hat{p}_z
-\hat{z}\hat{p}_y) + \frac{q^2}{4 m} B_0 xz \ ,
\end{eqnarray}
where a term $\hat{H}_2(t)$ of order $B_1^2$ has been neglected. Realistic
magnetic fields strengths impose a hierarchy between these terms because
usually $|B_0| \gg |B_1|$.

For comparison, two frequently used Hamiltonians which give the
regular Bloch equation, Eq.~(\ref{Blocheq}), are: $\hat{H}^{{\rm B}}=
\hat{H}_0^{ {\rm B}}+\hat{H}_1^{{\rm B}}(t)$, with:
\begin{eqnarray}
&& \hat{H}_0^{{\rm B}}=\hat{H}^{\prime}_0=\hat{V}(r)+\frac{\hat{p}^2}{2m}
-B_0 \hat{\mu}_z^{\prime}\ , \\
&& \hat{H}_1^{{\rm B}}(t) =\hat{H}^{\prime}_1 = - B_1\left(e^{i \omega
t}+e^{-i \omega t}\right) \hat{h}_1^{\prime}\ ; \ \
\hat{h}_1^{\prime}=\hat{ \mu}_x^{\prime}\ ,
\end{eqnarray}
obtained when all second order terms in the fields are neglected, as usual;
and the Hamiltonian for a pure spin system: 
\begin{eqnarray}
&& \hat{H}^{{\rm B}}_0 = g_s \mu_0 \hat{\vec{s}} \cdot\vec{B}_0 , \\
&& \hat{H}^{{\rm B}}_1 = g_s \mu_0 \hat{\vec{s}} \cdot\vec{B}_1 \left(e^{i
\omega t}+e^{-i\omega t}\right) \ .
\end{eqnarray}
In both these cases for which the Bloch equation applies, the
time-independent and time-dependent parts of the Hamiltonian commute:
$[\hat{ H}_0^{{\rm B}},\hat{H}_1^{{\rm B}}] = 0$, while
$[\hat{H}_0,\hat{H}_1] \neq 0 $.  Either way, the time-independent
Hamiltonian can be diagonalized in terms of a complete basis-set
$\{|m\rangle\}$:
\begin{eqnarray}
&&\hat{H}_0 |m\rangle= E_m |m\rangle \ ,
\end{eqnarray}
and the time-dependent electron wave function can be written (in the
Schr\"{o}dinger representation) as a linear combination of these basis-set
states: 
\begin{eqnarray}
|\Psi(t)\rangle= \sum_m c_m(t) e^{-i E_m t/\hbar} |m\rangle \ ,
\end{eqnarray}
with time-dependent coefficients, $c_m(t)$, that satisfy a set of coupled
equations, 
\begin{equation}
\frac{dc_m}{dt}= -\frac{i}{\hbar} \sum_l e^{-i\omega_{lm}t} \langle
m|\hat{H}_1|l\rangle c_l \ , \label{coupled}
\end{equation}
where $\omega_{lm} \equiv (E_l-E_m)/\hbar$. Power is absorbed by the system
at the following rate: 
\begin{eqnarray}
{\cal P}(\omega)=\left[\frac{dE}{dt}\right]_t&=& \left[ \frac{d}{dt} \sum_m
E_m |c_m|^2 \right]_t \\
&=& \sum_m \sum_l {i\omega_{lm}} c^*_m c_l e^{-i\omega_{lm}t} \langle
m|\hat{H}_1|l\rangle \label{pa0} \\
&=& \sum_m \sum_l c^*_m c_l e^{-i\omega_{lm}t} \left\langle m
\left|-\frac{i}{\hbar}[\hat{H}_0,\hat{H}_1]\right|l \right\rangle \ .
\end{eqnarray}

Three different cases can be considered:

\begin{itemize}
\item  Pure spin system with no coupling to the environment.

\item  Pure spin system with coupling to the environment.

\item  Diamagnetic system (magnetic field induced magnetic moment due to
orbital angular momentum) with no coupling to the environment.
\end{itemize}

In the first case of a pure spin system with no coupling to the environment,
as well as in the case of the approximate Hamiltonian leading to the regular
Bloch equation; $[\hat{H}_0^{{\rm B}},\hat{H}_1^{{\rm B}}] = 0$. As a
result, the power absorption is identically zero. In this case it would be
wrong to apply perturbation theory to the expansion of Eq.~(\ref{pa0}). This
can be understood on very general terms: a time-dependent perturbation which
commutes with the time-independent Hamiltonian can not induce transitions.

Different magnetic systems (including in particular pure spin systems) with
coupling to the environment were considered in the literature within two
basically different approaches. In the first approach, $T_1$ and $T_2$
processes are included when considering the dynamics, and power absorption
can thereby occur \cite{Textbooks,Kittel}. The $T_1$ and $T_2$ processes can
be described by relaxation terms added to the Bloch equation; the resulting
dynamics can no longer be described by a Hamiltonian.

In another approach introduced by Kubo and coworkers \cite{Kubo}, which we
would like to compare to, it is assumed that a Hamiltonian completely
defines the dynamics: 
\begin{eqnarray}
&&\hat{H}^{\prime\prime}=
\hat{H}_0^{\prime\prime}+\hat{H}_1^{\prime}(t) \ , \\
&&\hat{H}_1^{\prime}(t) = - B_1\left(e^{i \omega t}+e^{-i \omega
t}\right) \hat{h}_1^{\prime}\ ; \ \ \hat{h}_1^{\prime}\equiv
\mu^{\prime}_x= \frac{q}{2m}(\hat{y}\hat{p}_z -\hat{z}\hat{p}_y) \ .
\end{eqnarray}
$H_0^{\prime\prime}$ effectively incorporates the coupling to the
environment; although the coupling is not explicitly specified, it is
assumed that $[\hat{H}^{\prime\prime}_0,\hat{H}^{\prime}_1] \neq 0$. 
The power absorption therefore does not vanish, and can be calculated
using, for example, a linear-response theory \cite{Kubo}.

We now show that the power absorption of a
diamagnetic system with no coupling to the environment is very similar
to power absorption of a pure spin system with coupling to the
environment when the coupling to the environment is considered within
linear-response theory as in Ref.~\cite{Kubo}.  We show that in both
cases the power absorption does not vanish and is given by a similar
(though not identical) expression.  The similarities and differences
are derived and discussed below.  The essential point is that as soon
as $[\hat{H}_{0},\hat{H}_{1}]\neq 0$, power absorption does not
vanish, and, using the fact that for realistic magnetic fields
$B_{1}\ll B_{0}$, time-dependent perturbation theory can be applied.

The first-order solution of Eq.~(\ref{coupled}) with initial condition
at $t=0$, $c_{m}(0)=b_{m}$ is:
\begin{equation}
c_{m}=b_{m}+\frac{1}{\hbar }B_{1}\sum_{l}\langle
m|\hat{h}_{1}|l\rangle \left[ \frac{e^{i(\omega -\omega
_{lm})t}-1}{\omega -\omega _{lm}}-\frac{e^{-i(\omega -\omega
_{ml})t}-1}{\omega -\omega _{ml}}\right] b_{l}\ .
\end{equation}
Substituting this result into Eq.~(\ref{pa0}), and applying the rotating
wave approximation in the usual way, the power absorption is obtained: 
\begin{equation}
{\cal P}(\omega
)=2B_{1}^{2}\sum_{E_{n}>E_{m}}\frac{\omega_{nm}}{\hbar} |\langle
n|\hat{h}_{1}|m\rangle |^{2}\left( |b_{m}|^{2}-|b_{n}|^{2}\right)
\left[ \frac{\sin {(\omega -\omega _{nm})t}}{\omega -\omega
_{nm}}\right] _{t}\ .
\end{equation}
In the limit $t\rightarrow \infty $, we find: 
\begin{eqnarray}
&&{\cal P}(\omega )=2B_{1}^{2}\omega \chi _{x}^{\prime \prime }\ ,
\label{thepa} \\
&&\chi _{x}^{\prime \prime }=\pi \sum_{m,n}|\langle n|\hat{h}_{1}|m\rangle
|^{2}\left( |b_{m}|^{2}-|b_{n}|^{2}\right) \delta (E_{n}-E_{m}-\hbar \omega
)\ .  \label{sus}
\end{eqnarray}

In the case of a pure spin system with coupling to the environment
$\hat{h} _{1}=\hat{\mu}_{x}$.  Eq.~(\ref{sus}) gives the static
magnetic susceptibility and with Eq.~(\ref{thepa}) is identical to the
results obtained by Kubo and Nagamiya \cite{Kubo} in their Eqs.~(18.4)
and (19.17).  We note that they have obtained this result within a
different treatment, using a density matrix formalism, but under the
same assumptions as those supporting the perturbative treatment we
have used, namely, that the system is linear and in a steady-state. 
In the density matrix treatment, $|b_{m}|^{2}$ and $|b_{n}|^{2}$ in
Eq.~(\ref{sus}) are replaced by the densities $\rho (E_{m})$ and $\rho
(E_{n})$, respectively.

We note here that this expression for the power absorption is correct
only as long as a perturbative treatment is appropriate.  As soon as a
significant part of the population occupies the excited states a
nonperturbative treatment would be necessary.  The power absorption
would necessarily cease as the system continues to absorb energy from
the field and evolves to higher energy states if no coupling to the
environment were present, but this is not contained in a perturbation
theory or linear response approach.  Here we study the power
absorption only within the perturbative regime.  We note that the time
scale from turning on the fields till a significant change in the
population could be noticed can in principle be quite large.

In the case considered here, of no coupling to the environment but
inclusion of orbital effects, $\hat{h}_1 = \frac{q }{2m}
(\hat{y}\hat{p}_z -\hat{z} \hat{p}_y)+ \frac{q^2}{4 m} B_0
\hat{x}\hat{z}$, while in the case of coupling to the environment but
no orbital effects, $\hat{h}_1=\hat{\mu}_x^{\prime}= \frac{q }{2m}
(\hat{y}\hat{p}_z -\hat{z}\hat{p}_y) $.  Moreover, the spectrum and
hence the resonance frequencies are different in these two cases.  In
either case, once the Bloch equation is modified, whether by
relaxation terms or by terms of second order in the external fields,
power absorption can ensue.  Furthermore, while the numerical details
of this power absorption may differ, the general behavior is the same
in these two cases.  The power absorption is given by
Eq.~(\ref{thepa}) with a somewhat different quantum susceptibility in
the two cases.  In both cases, the quantum susceptibility is obtained
by substituting the appropriate energy eigenvalues and coupling
Hamiltonians into Eq.~(\ref{sus}).

To summarize this section: 
focusing on the regime where a linear response theory applies, we have
solved for the power absorption using time-dependent perturbation
theory.  We have derived Eqs.~(\ref{thepa}, \ref{sus}) for the power
absorption and for the susceptibility.  This result is similar to that obtained
by Kubo and coworkers in Ref.~\cite{Kubo} but there are several
differences: the coupling Hamiltonian $\hat{h}_{1}$ is different, the
quantum susceptibilities of the two systems are slightly different,
and the spectral position is shifted.  These are small corrections. 
Their order of magnitude is determined by the ratio between the
magnetic energy and the total energy of the system. Hence, it seems that
it would be difficult to experimentally observe a difference between the
result for the susceptibility obtained using our approach and the
standard result for  cw applications where a linear response theory
applies. The major difference is in justifying the linear response
treatment.  We have shown that it applies only if $[\hat{H}_0,\hat{H}_1]
\neq 0 $. This condition for a consistent application of perturbation
theory holds in Kubo's derivation only if coupling to the environment is
included, while for our case here no such coupling is required.

\section{Power absorption of an electron bound in a $p$-atomic-orbital}
\label{Example}

As an example, we study power absorption of an electron bound by the
Coulomb potential, $V(r)=-Z e^2 / (4\pi \epsilon_0 r)$, in a
$p$-orbital ($l=1$), and placed in the external time-dependent
magnetic field of Eq.~(\ref{B}).  Neglecting spin effects, the
complete Hamiltonian can be written as above.  As a basis set we
choose the three wave functions:
\begin{equation}
\langle\vec{r}|\psi_{\pm}\rangle = \frac{\pm 1}{\sqrt{2}}\ (x\pm iy)\
f(r) \ , \ \langle\vec{r}|\psi_{0}\rangle = z \ f(r) \ , \label{psi}
\end{equation}
where for principal quantum number $n = 2$, 
\begin{equation}
f(r) \equiv \frac{1}{4}\frac{1}{\sqrt{2\pi}} \left( \frac{Z}{a_0}
\right)^{5/2} \exp \left[ -\frac{Z}{2 a_0} \ r \right] \ ,
\end{equation}
and the Bohr radius is defined as $a_0 \equiv 4\pi \epsilon_0 \
\hbar^2/(m e^2)$.  These basis functions are eigenfunctions of the
free Hamiltonian when no magnetic fields are present, as well as of
$\hat{H}_0^{\prime}$, (but not of $\hat{H}$ or even $\hat{H}_0$).  The
$l=1$ manifold spanned by this basis set is not closed under the
complete Hamiltonian.  The orbital angular momentum $l$ is not a good
quantum number due to the terms in the Hamiltonian of second order in
the external magnetic field, which do not commute with the angular
momentum operator.  The coupling to other manifolds is small, however,
and the energy difference between the different manifolds is large. 
We therefore neglect the coupling to other manifolds, and consider the
action of $\hat{H}$ only within the $l=1$ subspace.  A general
solution of the time-dependent problem is then given by
\[
\ |\Psi(t) \rangle =c_{0}(t)e^{-iE_{0}t/\hbar }|\psi _{0}\rangle
+c_{+}(t)e^{-iE_{+}t/\hbar }|\psi _{+}\rangle +c_{-}(t)e^{-iE_{-}t/\hbar
}|\psi _{-}\rangle \ , 
\]
and the time-dependent perturbative analysis of the coefficients of
the previous section applies with $m=0,+,-$.

The diagonal matrix elements of $\hat{H}$ in this basis set are: 
\begin{eqnarray}
&&E_0\equiv \langle \psi_0|\hat{H}_0|\psi_0\rangle = - \frac{{\cal
E}}{8 }\ + \frac{12}{{\cal E}} (\mu_0 B_0 )^2\ , \\
&&E_\pm\equiv \langle \psi_\pm|\hat{H}_0|\psi_\pm\rangle = -
\frac{{\cal E}}{8}\ \pm \mu_0 B_0 + \frac{15}{{\cal E}} (\mu_0 B_0 )^2
\ ,
\end{eqnarray}
where ${\cal E} \equiv \frac{Z^2 e^2}{ 4\pi \epsilon_0 a_0}$. The level
spacings are given by, 
\begin{eqnarray}
\hbar\omega_+ \equiv E_+-E_0 = \mu_0 B_0 \left( 1+3\frac{\mu_0
B_0}{{\cal {E} }}\right) > 0 \ , \\
\hbar\omega_- \equiv E_0-E_- = \mu_0 B_0 \left( 1-3\frac{\mu_0
B_0}{{\cal {E} }}\right) > 0 \ ,
\end{eqnarray}
where $\hbar\omega_- > 0$ assuming $3\mu_0 B_0 < {\cal E}$.  The
oscillating field couples $|\psi_0\rangle$ and $|\psi_\pm\rangle$:
\begin{eqnarray}
&&\langle\psi_0|\hat{h}_1|\psi_\pm\rangle = \langle\psi_\pm|\hat{h}
_1|\psi_0\rangle = \frac{-\mu_0 }{\sqrt{2}}\ \left[1\mp 12 \frac{\mu_0
B_0}{{\cal E}} \right] \ .
\end{eqnarray}

The leading order contribution to the power absorption is: 
\begin{eqnarray}
{\cal P}_0(\omega) &=& \frac{\pi}{\hbar^2} \left(\mu_0 B_0\right)
\left(\mu_0 B_1\right)^2  \nonumber \\
&&\times \left\{ \left( |b_-|^2-|b_0|^2 \right)
\delta(\omega-\omega_-) + \left( |b_0|^2-|b_+|^2 \right)
\delta(\omega-\omega_+) \right\} .
\label{leadingorder}
\end{eqnarray}
The next order correction to this power absorption is: 
\begin{eqnarray}
{\cal P}_1(\omega) &=& \frac{24 \pi}{\hbar^2} \frac{(\mu_0 B_0)}{{\cal
{E}}} \left(\mu_0 B_0\right) \left(\mu_0 B_1\right)^2 \nonumber \\
&& \times \left\{ \left( |b_-|^2-|b_0|^2 \right)
\delta(\omega-\omega_-) + \left( |b_+|^2-|b_0|^2 \right)
\delta(\omega-\omega_+) \right\} .
\label{correction}
\end{eqnarray}
Higher order terms are of third order in $B_1$.  At temperature $T=0$,
$b_0=b_+=0$ and $b_-=1$, and we obtain:
\begin{eqnarray}
{\cal P}(\omega) &=& \frac{\pi}{\hbar^2} \left(\mu_0 B_0\right)
\left(\mu_0 B_1\right)^2 \left( 1+ 24 \frac{\mu_0 B_0}{{\cal E}}
\right) \delta\left[ \omega-\frac{\mu_0 B_0}{\hbar} \left( 1- 3
\frac{\mu_0 B_0}{{\cal E}} \right) \right] \ .  \label{final}
\end{eqnarray}
Note that for realistic experimental conditions, $\left(
\mu_{0}B_{1}\right) ^{2}$ is small enough to justify the perturbative
treatment and as a result ${\cal P}(\omega )$ is very small.

Realizable magnetic fields are, for example, $B_0 = 5$ T and
$B_1=5\times 10^{-4}$ T. The relative strength of the correction term
${\cal P}_1$ as compared to the leading order term ${\cal P}_0$, is
determined by the dimensionless parameter $24 \mu_0 B_0/ {\cal E}$,
which for $B_0=5$ T and $Z=6$ (carbon) is equal to $1 \times 10^{-5}$. 
${\cal P}_1$ as well as the higher order terms are thus but small
corrections to ${\cal P}_0(\omega)$.

\section{Conclusions and Summary} \label{Conclusions}

In this paper we have studied the dynamics of the gauge-invariant
operator for the magnetic dipole moment ${\vec{\mu}}$ of an electronic
system in an external magnetic field $\vec{B}$.  In addition to the
intrinsic magnetic moment of the electron it includes a field induced
magnetic moment arising from the electronic orbital angular momentum,
i.e., a diamagnetic contribution.  We have derived the Heisenberg
equation of motion for this magnetic moment in a general magnetic
field and shown that this modified Bloch equation is coupled to two
other Heisenberg equations for the position and velocity operators. 
While we could not solve this set of equations exactly and in closed
form, we have solved it perturbatively using time-dependent
perturbation theory and calculated the power absorption of the
magnetic moment in a time-dependent magnetic field under
magnetic-resonance conditions.  Under the assumption of a
steady-state, we have found an expression for the quantum
susceptibility in terms of matrix elements of a coupling Hamiltonian
$\hat{h}_{1}$.  Finally, we have explicitly calculated the integrated
power absorption for the example of a $p$-orbital electron and found
that it is given by Eq.~(\ref{final}).

Several generalizations can be easily considered. Upon generalizing to
include electron spin, the Hamiltonian and the magnetic moment operator
become, 
\begin{equation}
\hat{H} = \hat{V}(r) + \frac{\hat{\Pi}^{2}}{2m} -
\hat{\vec{\mu}}_{s}\cdot \vec{B} \ , \label{H_spin}
\end{equation}
\begin{equation}
\hat{\vec{\mu}}=\frac{q}{2m}\ (\hat{\vec{r}}\times \hat{\vec{\Pi}}
+g_{s}\hbar \hat{\vec{s}}) = \frac{-e}{2m}\ \hat{\vec{r}}\times
\hat{\vec{\Pi}} - g_{s}\mu _{0}\hat{\vec{s}}\ , \label{mu_spin}
\end{equation}
respectively, where $\hat{\vec{\mu}}_{s}=-g_{s}\mu _{0}\hat{\vec{s}}$,
and $g_{s}$ is the electron spin gyromagnetic ratio ($g_{s}\approx
2.002$).  The last term on the right hand side of Eq.~(\ref {mu_spin})
is the intrinsic spin magnetic moment operator of the electron,
$\hat{\vec{\mu}}_{s}$.  Incorporation of this term into the modified
Bloch equation yields the following term on the right hand side of
Eq.~(\ref{dmu} ): $-\frac{\mu _{0}g_{s}}{\hbar }\vec{B}\times
\hat{\vec{\mu}}_{s}$, i.e., the generalized Bloch equation becomes,
\begin{equation}
\frac{d\hat{\vec{\mu}}}{dt} = \frac{e^{2}}{4m}\left[ \left(
\hat{\vec{r}}\cdot \frac{\partial {\vec{B}}}{\partial t}\right) \
\hat{\vec{r}}\ -\ \hat{r^{2}} \frac{\partial {\vec{B}}}{\partial
t}\right] +\frac{e^{2}}{2m^{2}}\left[ \hat{\vec{r}}\times
(\vec{B}\times \hat{\vec{\Pi}})-i\hbar \vec{B}\right] + \frac{(\mu
_{0}g_{s})^{2}}{\hbar }\vec{B}\times \hat{\vec{s}}\ .
\label{dmus}
\end{equation}

Inclusion of spin-orbit coupling of the form $\alpha (r)\vec{s}\cdot
\vec{l}$ into the Hamiltonian is also straightforward.  The modified
Bloch equation then has additional terms on the right hand side
proportional to $\alpha (r) \vec{s}\times \vec{l}$ and $\alpha
(r)(\vec{s}\cdot \vec{r})(\vec{B}\cdot \vec{r})$.  The latter term is
due to the magnetically induced magnetic moment.  Furthermore, one can
also readily include ligand field effects, exchange coupling effects,
etc., within the formulation considered here.  Ligand field terms in
the Hamiltonian commute with the magnetically induced magnetic moment,
as do spin exchange terms, but orbital angular momentum exchange
coupling terms do not, and therefore contribute additional terms to
the extended Bloch equation due of the magnetically induced magnetic
moment.

How do our results differ from previous treatments of magnetic
resonance phenomena?  It is useful to divide such a comparison into
two subjects: the equations of motion and the mechanism of power
absorption.

We have shown that the Bloch equation, Eq.~(\ref{Blocheq}), is not the
correct Heisenberg equation for a magnetic moment $\vec{\mu}$ in the
magnetic field $\vec{B}$ when orbital angular momentum is present.  A
modified Bloch equation, Eq.~(\ref{dmu}), is obtained.  Upon inclusion
of spin, the factor of $g_{s}$ in the term $-\frac{\mu_{0}g_{s}}
{\hbar} \vec{B}\times \hat{\vec{\mu}}_{s}$, which is equivalent to the
last term on the right hand side of Eq.~(\ref{dmus}), further
precludes writing a Bloch equation of the form of Eq.~(\ref{Blocheq})
with $\hat{\vec{\mu}}$ replaced with $\hat{\vec{\mu}}_{l} +
\hat{\vec{\mu}}_{s}$.  The modified equation can not be written in
terms of $\vec{\mu}$ and $\vec{B}$ alone.  It contains the position
and kinetic velocity operators and therefore must be solved together
with dynamical equations for these operators.  In general,
$d\vec{\mu}/dt$ is not perpendicular to $\vec{B}$, and $\vec{\mu}$
oscillates out of phase with $\vec{B}$.  This results in a
non-vanishing power absorption.

In the usual picture, correct for pure spin systems, power
is absorbed only because of the coupling of the spin system to the
environment.  Without this coupling the magnetic moment $\vec{\mu}$
would oscillate in phase with $\vec{B}$, $d\vec{\mu}/dt$ would be
perpendicular to $\vec{B}$ at all times, and no power would be
absorbed.  Coupling to the environment dephases the oscillations of
$\vec{\mu}$ and $ \vec{B}$ and induces a nonvanishing component of
$d\vec{\mu}/dt$ along $\vec{B}$.  In the case considered here, of a
magnetic moment that explicitly contains a diamagnetic contribution,
the power absorption does not vanish even when coupling to the
environment is not included because the magnetic moment has a
component which oscillates out of phase with the external field as a
result of non-linear terms in the interaction Hamiltonian arising from
a magnetic field induced magnetic moment.  The phase difference
between oscillations of the magnetic field and oscillations of the
magnetic dipole moment of a bound electron in this field, is an
intrinsic feature of the system and need not be induced by additional
couplings to a bath that gives rise to dissipation.  This power
absorption can be interpreted, as evident from Eq.~(\ref{e_interp}), as
being due to the rate of change of the electric dipole moment in the
presence of an electric field.  Additional power aborption may result
due to coupling to a bath.

For cw power absorption, dissipation of energy from the atomic system
to the environment is required in order to ensure that the system
stays in thermal equilibrium, so that power absorption can be
maintained under steady-state conditions.  Hence, for truly cw power
absorption, as opposed to pulsed magnetic resonance experiments, the
two systems depend on coupling to the environment to maintain a steady
state and eventually will behave similarly.  The results presented
here indicate that diamagnetic systems would behave differently from
pure-spin systems during the initial time when the turning on of the
fields occur.  For pure spin systems the time scale for initiation of
power absorption would be set by the coupling to the environment while
for diamagnetic systems it would be an intrinsic time scale
indifferent to the environment.  Without a relaxation mechanism, no
power would be absorbed by a pure spin system, while the diamagnetic
system will absorb power upon the turning on of the fields only until
the population of levels reach a new equilibrium.  We note that as
soon as the population is changed significantly, second and higher
order effects become important and a first-order time-dependent
perturbation theory would not suffice.  

\begin{acknowledgements}
Useful discussions with Garnett Bryantt are gratefully acknowledged. 
This research was supported by The Israel Science Foundation (B.\ S.
is supported by grant No.~181/00 and Y.\ B.\ B. by grant No.~212/01).
\end{acknowledgements}

\end{document}